\documentstyle[proceedings,psfig]{crckapb}

\def\ni{\noindent}
\font\syvec=cmbsy10                        
\font\gkvec=cmmib10                         
\def\bnabla{\hbox{{\syvec\char114}}}       
\def\bxi{\hbox{{\gkvec\char24}}}           
\def\bcdot{{\bf \cdot}}
\def\bv{{\bf v}}

\def\bF{{\bf F}}
\def\bg{{\bf g}}
\def\dE_a{{\cal P}_\alpha}

\def\cC{{\cal C}}
\newcount\eqnumber
\def\chaphead{}
\def\last{\advance\eqnumber by -1 {\rm\chaphead\the\eqnumber}\advance
     \eqnumber by 1}
\def\ref#1{\advance\eqnumber by -#1 \chaphead\the\eqnumber
     \advance\eqnumber by #1 }
\def\new{{\rm\chaphead\the\eqnumber}\global\advance\eqnumber by 1}

\def\refind{\noindent \hangindent=2pc \hangafter=1}
\def\spose#1{\hbox to 0pt{#1\hss}}
\def\gta{\mathrel{\spose{\lower 3pt\hbox{$\mathchar"218$}}
     \raise 2.0pt\hbox{$\mathchar"13E$}}}
\def\lta{\mathrel{\spose{\lower 3pt\hbox{$\mathchar"218$}}
     \raise 2.0pt\hbox{$\mathchar"13C$}}}

\begin{opening}
\title{Excitation of Solar Acoustic Oscillations}

\author{Pawan Kumar}

\institute{Institute for Advanced Study, Princeton, NJ 08540}
\end{opening}

\begin{document}
\vglue-.86in
\hskip1.7in\hbox{$^1$}
\vglue.76in
\section*{Abstract}

The stochastic excitation of solar oscillations due 
\footnotetext[1]{Alfred P. Sloan Fellow \& NSF Young Investigator}
to turbulent convection is reviewed. A number of different observational
results that provide test for solar p-mode excitation theories are described.
I discuss how well the stochastic excitation theory does in explaining
these observations. The location and properties of sources that excite 
solar p-modes are also described. Finally, I discuss why solar g-modes 
should be linearly stable, and estimate the surface velocity amplitudes
of low degree g-modes assuming that they are stochastically excited
by the turbulent convection in the sun.

\setcounter{footnote}{1}

\section{Introduction}

It was realized about 25 years ago that the Sun, our nearest star, is
a variable star. Millions of acoustic normal modes (p-modes) of the sun 
are seen to be excited with a typical surface velocity amplitude 
of only a few cm s$^{-1}$, whereas other pulsating stars have a 
few modes excited to large amplitudes. 
Considering this dramatic difference between the pulsation property of the
sun and other variable stars, it should not be surprising that the
solar oscillations are excited by a mechanism that is different from the
overstability mechanism believed to be responsible for the pulsation of other 
stars (overstability can arise for instance when the radiative
flux is converted to mechanical energy of pulsation due to an increase of
opacity with temperature). A number of early papers in the field proposed 
that the solar p-modes are excited by some overstability mechanism
(Ulrich 1970, Leibacher and Stein 1971, Wolf 1972, Ando and Osaki 1975).
However, the margin of instability for solar p-modes is found to be small 
and different ways of handling radiative transfer and/or the interaction of 
convection with oscillation seems to change the sign of stability 
e.g. Goldreich and Keeley 1977a, Antia et al. 1982 \& 1988, 
Christensen-Dalsgaard and Frandsen 1983, Balmforth \& Gough 1990, 
Balmforth 1992 (the last two papers used a sophisticated version of the
mixing length theory of Gough, 1977); see Cox et al. 1991 for a more 
complete list of references.
If we assume that the solar p-modes are overstable then their
amplitudes grow exponentially until some nonlinear mechanism limits their
growth. By considering all possible 3-mode nonlinear couplings amongst 
overstable and stable p-modes in the sun, which is the most efficient 
process for saturating the amplitudes of overstable modes, 
Kumar and Goldreich (1989) and Kumar, Goldreich and Kerswell (1991) 
showed that the amplitudes of overstable modes saturate at a value 
that is several orders of magnitude larger than the observed value. 
This suggests that solar p-modes are linearly stable.

In this article we will assume that solar p-modes are stable, and describe
how they can be excited by acoustic waves generated by turbulent convection. 
The basic idea is that the broad band acoustic noise generated by the turbulent
flow in the convection zone is selectively amplified at frequencies 
corresponding to the normal mode frequencies of the sun. The process 
of wave generation by homogeneous turbulence was first studied systematically
and in some detail by Lighthill (1952). Stein (1967) and Kulsrud (1955) applied
it to the heating of the solar chromosphere/corona by acoustic and MHD waves
respectively. Goldreich and Keeley (1977b) carried out a careful calculation
of the stochastic excitation of solar normal modes by turbulent convection
(for an excellent general review of wave generation due to turbulent
fluid please see Crighton 1975).
We describe the stochastic excitation process for the simple case of
a homogeneous sphere below and later discuss its generalization to the Sun
(\S 2). In \S 3 we describe various observations that any theory for
excitation of solar p-modes must be able to explain and discuss how well the 
stochastic excitation theory performs when confronted with these observations.
The estimate for the surface velocity amplitude of low degree g-modes, assuming
that they are stochastically excited, is given in \S4.

\section{Stochastic Excitation}

Let us consider a homogeneous gas sphere with a surface that reflects 
acoustic waves. Some fraction of the fluid inside this sphere is assumed 
to be in the state of turbulence which acts as a source of sound waves. 
Following Lighthill (1952) we write the perturbed mass and momentum 
equations as

$$ \rho_1 + \bnabla\cdot(\rho{\bf \bxi}) = 0, \eqno(\new)$$
\ni and
$$ {\partial^2\rho\xi_i\over\partial t^2}+c^2{\partial\rho_1\over\partial x_i}
    = -{\partial T_{ij}\over \partial x_j}, \eqno(\new)$$
\ni where $c$ and $\rho$ are unperturbed mean sound speed and density of 
the medium, $\bxi$ is fluid displacement, and
$$ T_{ij} \equiv \rho v_i v_j + p\delta_{ij} - \rho c^2 \delta_{ij}. 
\eqno(\new)$$

\ni These equations can be combined to yield the following 
inhomogeneous wave equation 
$$ {\partial^2\rho \xi_i\over \partial t^2} - c^2 \nabla^2 (\rho \xi_i)
    = -{\partial T_{ij}\over \partial x_j}, \eqno(\new)$$

\ni Expanding ${\bxi}$ in the terms of normal modes of the system

$$ {\bxi} = {1\over\sqrt{2}}\sum_q A_q {\bxi}_q \exp(-i\omega t) + c.c., 
\eqno(\new)$$
where ${\bxi}_q$ is displacement eigenfunction of mode $q$ which is
normalized to unit energy {\it i.e.}
$$ \omega_q^2\int d^3x\, \rho\,{\bxi}_q\cdot{\bxi}_q^* = 1, \eqno(\new)$$

\ni and substituting this expansion into equation (\ref3) we find the following
equation for the mode amplitude $A_q$

$$ {d A_q\over dt} \approx -{i\omega_q\over\sqrt{2}}\exp(i\omega_q t)\int 
   d^3x\,\xi_{q_{i}} {\partial T_{ij}\over \partial x_j} = 
   {i\omega_q\over\sqrt{2}}\exp(i\omega_q t)\int d^3x\, 
    {\partial \xi_{q_{i}} \over\partial x_j} T_{ij}. \eqno(\new)$$
    
\ni Turbulent flow is crudely described as consisting of critically damped 
eddies. The velocity $v_h$ of an eddy of size $h$ is related to the largest 
or the energy bearing eddy (size $H$ and velocity $v_H$) by the Kolmogorov 
scaling {\it i.e.}
$$ v_h = v_H \left( {h\over H}\right)^{1/3}. \eqno(\new)$$
Moreover, following Lighthill (1952), we take $T_{ij}\approx 
\rho v^2\delta_{ij}$. Since the displacement eigenfunction, for low $\ell$ 
modes, near the surface of the sphere is in the radial direction, therefore 
equation (\ref2) reduces to

$$ {d A_q\over dt} \approx {i\omega_q\over\sqrt{2}}\exp(i\omega_q t)
   \int d^3x\, \rho v^2 {\partial \xi_{q_{r}} \over \partial r}. \eqno(\new)$$
where $\xi_{q_{r}}$ is the radial displacement eigenfunction of mode $q$.
The mean energy input rate in mode $q$ can be obtained from
the above equation and is given by

$$ {d E_q\over dt} \equiv {d \langle |A_q|^2\rangle\over dt}\approx 
   2\pi\omega_q^2 \int dr\; r^2\,\rho^2 v_\omega^3 h_\omega^4 \left[{\partial 
   \xi_{q_{r}}\over\partial r}\right]^2, \eqno(\new) $$ 
\ni where $v_\omega$ and $h_\omega$ are the velocity and size of the
eddies which have characteristic time, $\tau_h\equiv h_\omega/v_\omega$, 
approximately equal to the mode period. This
equation is valid not only for the homogeneous gas sphere considered here but 
also for more general systems including the excitation of solar p-modes
by the Reynolds stress as discussed below.
Of course, we must use the eigenfunction $\xi_q$ and turbulent 
velocity appropriate for the system being considered.

It can be easily shown that the solution of the homogeneous wave equation
(eq. [4] with right side set equal to zero), in the limit of large $n$ (mode 
order) is 

$$\xi_{qr} \approx B \; j_\ell\left(\omega_q r/c\right)
      \approx B {\sin(r\omega_q/c - \pi\ell/2)\over r\omega_q}, \eqno(\new)$$
where $j_\ell$ is spherical Bessel function, and $B$ is a constant factor
independent of mode frequency for properly normalized mode eigenfunction 
(condition expressed by eq. [6]).
Substituting this into equation (\ref2) we find

$$ \dot E_q\equiv {d E_q\over dt} \propto \omega_q^2 h_\omega^4 v_\omega^4. 
\eqno(\new)$$

Let us assume that the turbulent velocity field in the sphere is concentrated 
in a thin layer of thickness $H$ located near the surface of the sphere.
We shall take the size of the largest eddies to be $H$ and their rms speed to 
be $v_H$. The p-modes of period greater than $\tau_H= H/v_H$ are 
predominantly excited by the largest size eddies and the resultant 
energy input rate in these modes is proportional to $\omega_q^2$, 
as can be seen immediately from the above equation. Modes of higher frequency
($\omega_q\gta\tau_H^{-1}$) couple best to inertial range eddies which have 
characteristic time of order the wave period. Making use of the Kolmogorov
scaling (eq. [8]) and equation (\last) we see that $\dot E_q$ 
scales as $\omega_q^{-5.5}$. Thus the energy input rate into p-modes of this 
homogeneous system shows a break at frequency $1/\tau_H$ where the
power law index changes by 7.5.

The generalization of above equations to describe the excitation of solar
modes is not difficult. Equation (2) is replaced by the linearized momentum 
equation valid for a stratified medium i.e. 

$$ \rho {\partial^2\bxi\over\partial t^2} + \bnabla p_1
-\rho_1\bg = -\bnabla\bcdot(\rho\bv\bv) \equiv \bF, \eqno(\new)$$
and the linerized equation of state is
$$p_1 = {\partial p\over \partial \rho}\rho_1 + {\partial p\over
\partial s} s_1, \eqno(\new)$$
where
$$ s_1 = \tilde{s} - (\bxi\bcdot\bnabla) s. \eqno(\new)$$
\ni Here $\bnabla s$ denotes the background entropy gradient, and
$\tilde{s}$ is the entropy fluctuation associated with turbulent
convection. Equation (\last) is the Eulerian version of the
statement that the Lagrangian entropy perturbation is due
entirely to turbulent convection. In other words, we approximate
the waves as adiabatic (these equations are adopted directly from 
Goldreich et al. 1994). Combining equations (1) and (\ref3)-(\ref1) we obtain
the following inhomogeneous wave equation, which is the generalization of 
equation (4) and describes the stochastic excitation of solar oscillations:
$$ \rho {\partial^2\bxi\over\partial t^2}-\bnabla\left[c^2\bnabla
\bcdot\left(\rho\bxi\right)+\rho\bxi\bcdot\bg-c^2\rho\bxi\bcdot
\bnabla\ln\rho\right]+\bg\bnabla\bcdot\left(\rho\bxi\right)
 = - \bnabla\left({\partial p\over\partial
s}\tilde{s}\right)-\bnabla\bcdot\left(\rho\bv\bv\right). \eqno(\new)$$
\ni This equation describes wave generation due to Reynolds stress
as well as entropy fluctuation.
As the entropy of a fluid element fluctuates, so does its volume.
The fluctuating volume is a monopole source for acoustic waves.
In a stratified medium the fluctuating buoyancy force adds a
dipole source.  By transferring momentum among neighboring fluid
elements, the Reynolds stress acts as a quadrupole source.\footnote{We
classify acoustic sources as monopole, dipole, or quadrupole
according to whether they produce a change in volume, add net
momentum, or merely redistribute momentum.}~ The anisotropy of a
stratified medium blurs the distinction between monopole, dipole,
and quadrupole sources. It allows for destructive interference
between the monopole and dipole amplitudes. Although the monopole
and dipole amplitudes are individually larger than the quadrupole
amplitude, their sum is of comparable size to that of the
quadrupole. That this applies to energy bearing eddies follows
directly from equations (\last) and the relation between entropy and
velocity fluctuation for convective eddies. The justification for inertial 
range eddies requires a subtle argument (cf. Goldreich and Kumar, 1990).

\begin{figure}[ht]
\centerline{\psfig{figure=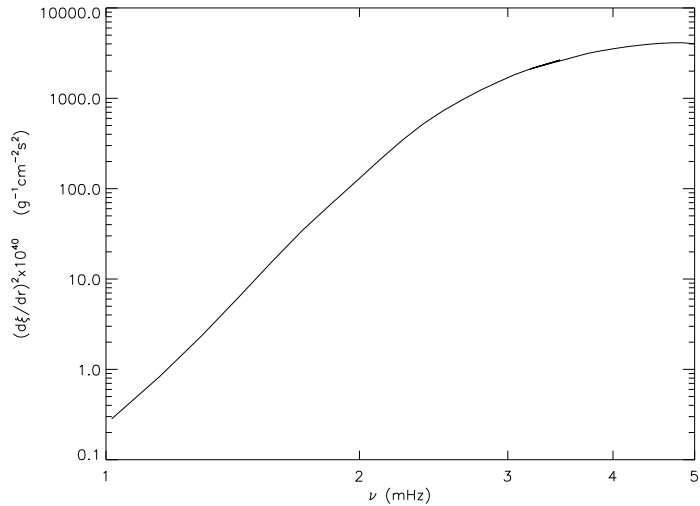,width=3.7in,angle=0.0}}
\caption[]{The plot of $(d\xi_r/dr)^2$, at the top of the solar convection 
zone,
as a function of p-mode frequency (the mode degree is zero). The solar 
model used here is due to J. Christensen-Dalsgaard.
}
\end{figure}

The new energy equation (which replaces eq. [10]) is given below

$$ \dot E_\alpha\sim 2\pi \omega_\alpha^2\int\, dr\, r^2\, 
\rho^2\left|{\partial
  \xi_\alpha^r \over \partial r}\right|^2\, v_\omega^3h_\omega^4 
  \left(\cC_\alpha^2{\cal R}^2 +1\right){\cal S}^2, \eqno(\new)$$
where $\cC_\alpha$ is wave compressibility defined by
$$ \bnabla\bcdot\bxi_\alpha = \cC_\alpha {\partial\xi_\alpha^r\over\partial r},
   \eqno(\new)$$
the shape parameter ${\cal S}$ describes the ratio of the
horizontal to vertical correlation lengths of turbulent eddies, and ${\cal R}$
is given by
$$ {\cal R}\equiv  {4H\over \Lambda} \left({\partial\ln p\over \partial\ln
\rho}\right)_s, \eqno(\new)$$
with $\Lambda$ the mixing length and $H$ the pressure scale height.
The factor $\cC_\alpha^2{\cal R}^2$ measures the ratio of the excitation by 
entropy fluctuations to that due to fluctuating Reynolds stress. 
Note that the frequency spectra of waves excited due to entropy fluctuation 
and that due to Reynolds stress are identical except of course for an
over all normalization factor. The compressibility, 
$\cC_\alpha$, for p-modes near the top of the convection zone, where 
the excitation takes place, is close to 1, and the value of 
${\cal R}^2$ in this region is about 10 (see Goldreich et al. 1994). 
Therefore, the excitation of p-modes is dominated by entropy fluctuations 
(Stein \& Nordlund 1991, arrived at the same conclusion using their 
numerical simulations of solar convection). On the other hand f-modes are 
nearly incompressible ($\cC_\alpha\approx 0$) and so they are not excited 
by entropy fluctuation. This is perhaps why the power in f-modes is
observed to be smaller than a p-mode of similar frequency. 

The observed rate of energy input into solar p-modes can now be readily 
understood. One of the differences between the homogeneous gas sphere 
system and the sun is in the shape of the eigenfunction
especially near the surface where the wave excitation takes place. 
It can be shown that the radial derivative of the normalized radial 
displacement eigenfunction for p-modes just below the photosphere scales 
as $\nu_q^{3.8}$ for $\nu_q\lta 3.0$mHz and as $\nu_q^{1.1}$ for 
$\nu_q\gta 3.5$mHz (see Figure 1). Substituting this scaling 
in equation (\ref3), or equation (10), we find that the energy input rate 
in the p-modes, at a fixed
degree, scales as $\nu_q^7$ for $\nu_q\lta 3.0$mHz and as $\nu_q^{-4.4}$
for $\nu_q\gta 3.5$mHz which is in good agreement with the observations
(Libbrecht \& Woodard 1991); please see Goldreich et al. (1994) for a 
more detailed analysis and comparison with the observed energy input rate.

\section{Observational constraints for the excitation theory of
solar p-modes}

A valid theory for the excitation of solar p-modes should be able to explain 
the observed rate of energy input in different modes. In addition, 
there are four other observational results that the theory must be 
able to explain and provide a fit to the data. These observations are:
mode linewidth, the deviation of p-mode line profiles from symmetric 
Lorentzian shape, the statistics for the fluctuation of mode energy, and 
the presence of peaks in the power spectrum above the acoustic cutoff 
frequency ($\nu\gta 5.3$mHz).

The agreement between the observed and the theoretically calculated energy 
input rate in solar p-modes due to stochastic excitation was described in 
the last section. We describe below the other four observations and compare 
them with the results of the stochastic excitation theory.

\subsection{Mode Linewidth}

A number of groups have measured p-mode linewidth as a function of mode 
frequency (cf. Duvall et al. 1988, Libbrecht 1988, Elsworth et al. 1990). 
It is found that the mode linewidth at a fixed degree increases 
monotonically with mode frequency. At 2 mHz the linewidth is about 
0.5 $\mu$Hz, or mode lifetime is 20 days ($Q\approx 4\times10^3$) and at 
4 mHz the linewidth is 10 $\mu$Hz. The observed linewidth for 2mHz$\lta\nu
\lta 3$mHz increases as $\nu^{4.2}$ whereas numerical calculations 
show that the mode linewidth due to radiative and turbulent damping increases 
as $\nu^8$ for frequencies below $\sim$4mHz (cf. Christensen-Dalsgaard 
\& Frandsen 1983, Balmforth 1992; Goldreich and Kumar 1991, give a simple
analytical derivation of these results). This suggests that the mode damping
at frequencies greater than about 2 mHz is due to some process other than 
the radiative and turbulent viscosity. A number of alternate
mechanisms have been suggested to account for the observed mode linewidth.
These include modulation of convective flux (Christensen-Dalsgaard et al. 
1989), scattering of low degree p-modes by turbulent convection
to high degree modes (Goldreich and Murray 1994), scattering of p-modes 
by magnetic fields (Bogdan et al. 1996). Goldreich and Murray (1994) have 
carried out a detailed calculation of the scattering process and find that
an almost elastic scattering of p-modes by convective eddies is an 
important contributor to the mode linewidth at frequencies $\nu\gta 2$mHz,
and the computed linewidth has the same frequency dependence as the observed
width (see also Murray, 1993).

\begin{figure}[b]
\centerline{\psfig{figure=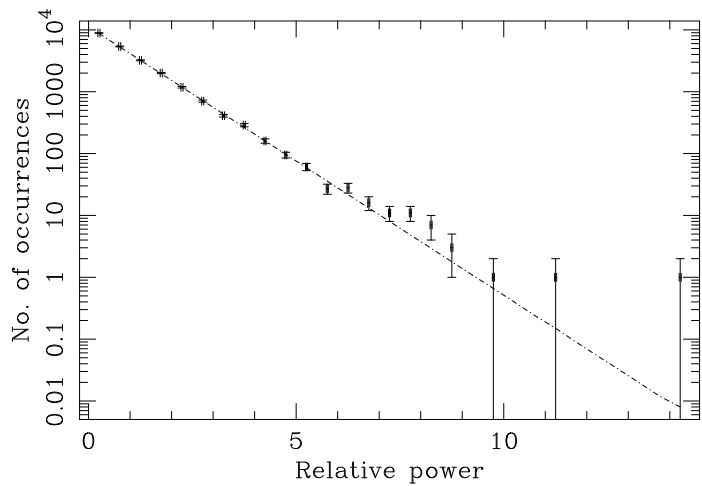,width=4in,angle=0.0}}
\caption[]{The statistics of power fluctuation in low degree p-modes.
The straight line is the exponential distribution, which is the theoretically
expected distribution if modes are stochastically excited due to turbulent 
convection. The data is kindly provided by BiSON (please see 
Chaplin et al. 1995 for details).
}
\end{figure}

Recent observational results indicate that the linewidth scales as 
$\nu^7$ for $\nu\lta 2$ mHz (Chaplin et al. 1996, and Tomczyk 1996).
Perhaps below 2 mHz there are few modes available for p-modes to scatter
into, and thus the linewidth falls off more rapidly with decreasing 
frequency. According to Jefferies (personal communication) the 
observed mode linewidth peaks at a frequency of about 5mHz, followed by 
a slight decline, and then remains constant at higher frequencies. 
This is a puzzling result for which as far as I know no explanation has 
been offered.

\subsection{Energy Statistics}

Modes excited by their interaction with a Gaussian random field (turbulent
convection) have fluctuating amplitudes that follows the Gaussian distribution.
The correlation time for mode amplitude, which is of order the mode 
lifetime, is typically much larger than the mode period (see \S 3.1) or
the characteristic time of resonant eddies.
This is because a mode interacts with a large number of eddies each of 
which contribute only a small fraction of the total energy in the mode. 
A good analogy is a pendulum placed in contact with a thermal heat bath
of molecules. The mean energy in the pendulum is one third the mean 
kinetic energy of molecules, however it takes a large number of collisions 
(of order the ratio of the pendulum mass to molecular mass) in order for 
the amplitude of the pendulum to change. The statistics of energy 
fluctuation in a solar p-mode, if stochastically excited, like the energy 
of the pendulum placed in a heat bath, follows Boltzmann distribution
(see Kumar et al. 1988 for a rigorous derivation of this result).
At least two different groups (Toutain \& Frohlich 1992; and Elsworth et al.
1995) have looked for the statistics of energy fluctuation in the solar
p-modes and find it to be in good agreement with the theoretical expectation
for stochastic excitation i.e. Boltzmann or exponential distribution (see
fig. 2).

\subsection{Peaks at high frequencies}

\begin{figure}[ht]
\centerline{\psfig{figure=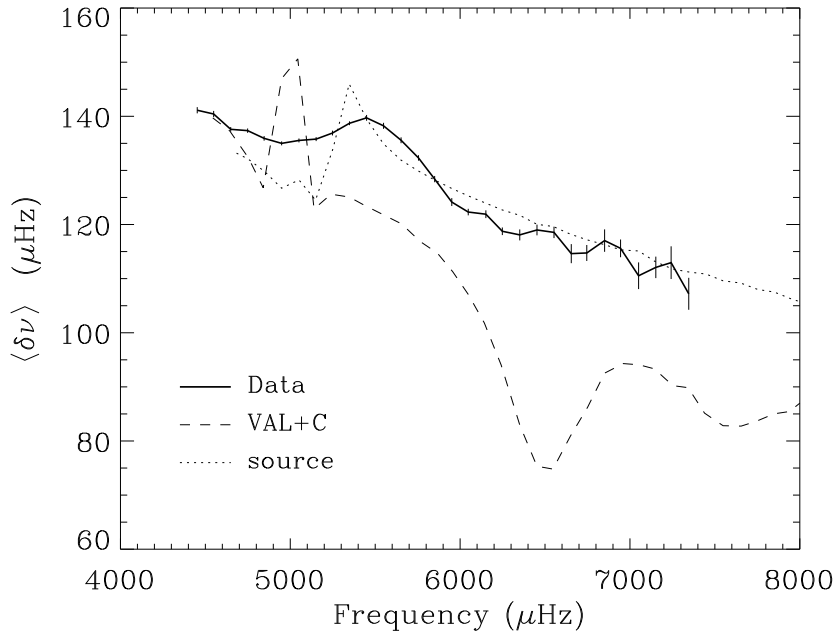,width=4.7in,angle=0.0}}
\caption[]{Average frequency spacing between adjacent 
peaks in the power spectrum,
$\langle\delta\nu\rangle$, as a function of frequency. The averaging over
frequency bins of width 100 $\mu$Hz, and $\ell$ range of 80 and 150
has been carried out after subtracting a linear term in $\ell$ 
(0.6981 $\ell$ $\mu$Hz) from $(\nu_{n+1,\ell} - \nu_{n,\ell})$.
The observational data (thick solid curve)
was obtained by Duvall et al. (1993) at the geographical South Pole in
1988. The dotted curve, labled `source' in the legend, is the result of 
calculation of peak frequencies in the theoretically computed power spectra 
for Christensen-Dalsgaard's solar model with sources lying about 140 km below
the photosphere. The dashed curve labled `VAL+C' is the frequency spacing 
calculated for JC-D solar model that includes the ``mean quiet sun'' 
chromospheric structure of Vernazza et al. (1981) as well as an isothermal 
corona at a temperature of 10$^6$ K.
}
\end{figure}

Acoustic waves of frequency less than about 5 mHz (the acoustic cutoff
frequency at the temperature minimum) are reflected 
at the solar photosphere and thus trapped inside the sun. The reflectivity 
however drops off rapidly at higher 
frequencies; at 6mHz about 2\% of the incident wave energy is reflected at the
photosphere whereas at 7mHz the reflectivity drops to less than 0.3\%. A
number of observations indicate that high frequency acoustic waves (waves
of frequency greater than about 5mHz) suffer little reflection at the 
chromosphere/corona as well (Duvall et al. 1993, Kumar et al. 
1994, Jefferies 1996). If high frequency acoustic waves were significantly 
reflected at the chromosphere/corona boundary then the frequency spacing 
between modes of adjacent order would fluctuate with mode frequency 
(see figure 3); this is because of the interference of waves partially 
trapped between two cavities above and below the temperature minimum. 
The observations, however, show, no evidence for such a behavior (fig. 3) 
and thus provide an upper limit of about 10\% to the reflection at
the chromosphere/corona boundary (Kumar et al. 1991; Jefferies, personal
communication). 

In the absence of wave reflection at the solar surface these high frequency 
acoustic waves are not trapped in the sun, and thus it was expected that the 
power spectrum above the acoustic cutoff frequency should be featureless
i.e. devoid of peaks. However, the 
observed spectra contain very regular peaks that are seen upto the
Nyquist frequency of observations. One of the best recent data 
set obtained at the South Pole in 1994 shows peaks extending to
almost 11 mHz which makes the length of the spectrum above the acoustic
cutoff frequency larger than the observed spectrum below the cutoff frequency!

The existence of these high frequency peaks provides one of the strongest
evidence that solar acoustic oscillations above 5 mHz are not excited 
by some overstability mechanism\footnote{Considering the poor reflectivity
of high frequency waves at the chromosphere/corona, the energy flux in the
solar atmosphere associated with them represents a net loss of their energy.
So if these waves are to be excited due to 
an overstability mechanism, their $e$-folding time must be less than
about an hour, and thus these waves can at best be amplified by a factor
of $\sim e$ as they make one passage through the solar interior. Thus we need 
a mechanism that provides a large seed amplitude, within a factor of $e$ of 
the observed value, and clearly in this case it seems most natural that the 
same mechanism generates the full observed amplitude.}, 
and since power spectrum varies smoothly from below the acoustic cutoff 
frequency to above the cutoff frequency
we infer that the trapped p-modes in the sun are also not overstable 
(Kumar et al. 1989).

The peaks at high frequencies can be understood as arising quite naturally
if waves are stochastically excited. These peaks form because of the 
constructive interference between waves propagating from the source (located 
in the convection zone) upward to the photosphere and waves traveling downward 
from the source that is refracted back up due to increasing sound speed and
thus end up at the photosphere (Kumar and Lu 1991).
Therefore the frequencies of peaks above the acoustic cutoff ($\sim 5$ mHz for 
the sun) depend on the difference between these two paths or in other 
words on the depth of acoustic sources. A good fit to the high frequency 
power spectrum is obtained by placing sources (assumed to be quadrupole) 
approximately 140 km below the photosphere (Kumar 1994); please see 
figure 4. It should be emphasized that unlike
the lower frequency trapped p-modes (frequency less than about 5 mHz)
the frequencies of peaks at high frequencies is not a property
of the equilibrium model of the sun alone but depends in a sensitive way on
the location of sources that excite these oscillations.

\begin{figure}[ht]
\centerline{\psfig{figure=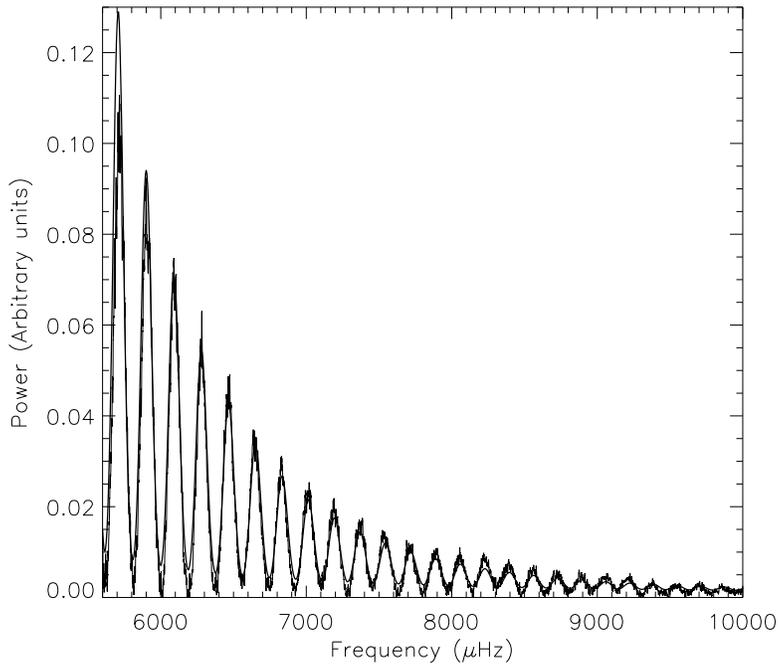,width=4in,angle=0.0}}
\caption[]{Observed power spectrum (thin solid line) from the 1994 
South Pole observation (courtesy of S. Jefferies) for $\ell$=90, and 
theoretically calculated power spectrum for sources lying 140km below (thick 
solid line) the photosphere. The Nyquist frequency of the data is 11.9 mHz.
Both the $\ell$--leakage and the Nyquist folding 
has been included in the theoretically computed spectrum; to model the 
$\ell$--leakage the theoretically calculated power spectra for $\ell$=88 to 
92 were added together with weighting factors of 0.147, 0.68, 1.0, 0.61, and
0.10 respectively which corresponds to the 1994 South Pole observations 
(Jefferies, personal communication). The radial extent of the sources is 
taken to be 50km, and the spectrum of turbulent convection is Kolmogorov. 
A frequency dependent background has been subtracted from the observed 
spectrum.
}
\end{figure}

As discussed in Kumar (1994) if the acoustic sources are assumed to be dipolar 
instead, then no matter where these sources are placed in the solar convection
zone they do not provide a fit to the observed power spectrum. This
suggests that the acoustic sources, at least for the high frequency waves,
are not dipole but quadrupole, which is consistent with the work of
Goldreich et al. (1994).

High frequency acoustic waves also provide information about the power
spectrum of turbulent convection in the sun (Kumar 1994); we can constrain
the spectrum of turbulent convection in the region where acoustic emission
is significant. The theoretical power spectrum
shown in figure 4 was computed using the Kolmogorov power spectrum of
turbulence, {\it i.e.}, $P(k)\propto k^{-5/3}$. Evidently, this provides a
good fit to the observed spectrum between 5.5 and 10 mHz.
In order to determine the power law index $\alpha$ of solar turbulence,
$P(k)\propto k^{-\alpha}$, from the high frequency interference peaks, 
we relate the fluctuating velocity, $v_h$, of sub-energy bearing eddies
to the velocity $v_H$ of scale-height size eddies as follows:
$$ v_h \approx v_H\left({h\over H}\right)^{(\alpha-1)/2}, $$
where $h$ is the size of the eddy. This equation is a  generalization of
equation (8). Using this relation, we find that the frequency
dependence of the source function is $(\omega\tau_H)^{-(3\alpha+5)/(3-\alpha)}$
(the derivation is similar to the one leading to eq. [12]). Therefore,
a change in the spectral index of turbulence from 5/3 to 1.4
decreases the dependence of the acoustic power spectrum on frequency by
$\omega^{1.75}$, which results in a poor fit to the observed spectrum.
We find that the observed high frequency power spectra suggest 
that the power law index for the solar turbulence lies between 1.5 and 1.7.

We note that the energy input rate for p--modes in the frequency range between
3.5 and 5mHz is proportional to $\omega^{-4.4}$, which is understood
most naturally if the spectrum of turbulence near the top of the
solar convection zone is taken to be Kolmogorov (Goldreich {et al. 1993}).
The result described above extends this range to 10mHz.

\subsection{Asymmetric line profiles of p-modes}

\begin{figure}[b]
\centerline{\psfig{figure=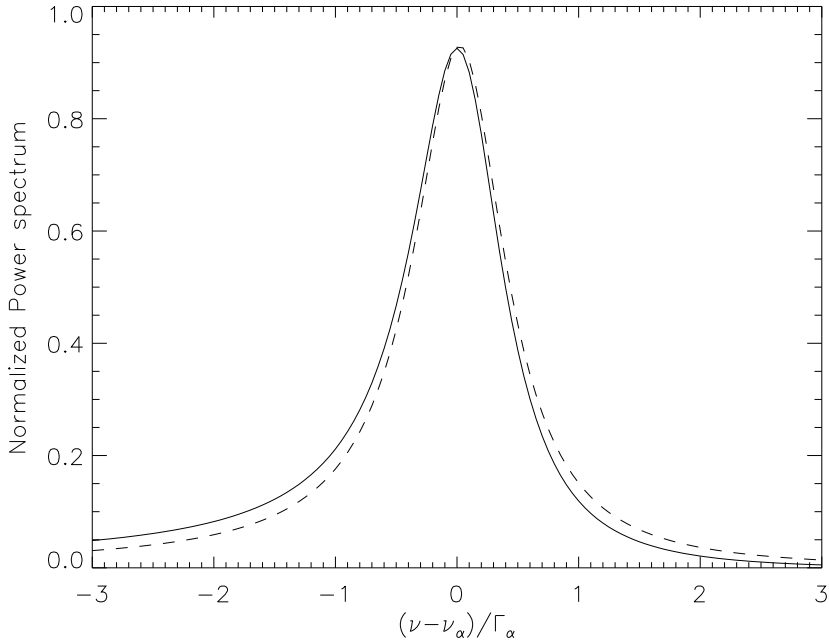,width=4.8in,angle=0.0}}
\caption[]{Line profiles of a p-mode of frequency 1.9 mHz and $\ell=1$
of a solar model due to J. Christensen-Dalsgaard. The power spectrum 
plotted using a continuous line corresponds to sources placed
at the upper turning point of the mode. The other power spectrum, dashed curve,
arises when sources are placed 300 km below the upper turning point.
The radial extent of sources in both cases was taken to be 100 km.
}
\end{figure}

The last observational result I would like to describe is the asymmetry of
low frequency p-mode line
profiles. The spectrum of individual p-modes is fitted very well by
a Lorentzian profile. However, Duvall et al. (1993) discovered that the
line profiles do not have perfect Lorentzian shape and in particular
the power spectrum falls off more rapidly on one side of the peak than the
other i.e. the lines are asymmetrical. The data from GONG and SOHO
clearly show that lineprofiles for low frequency modes are asymmetrical.
Duvall et al. had also proposed in their original paper
an explanation for why the lines are asymmetrical which is found to be
basically correct by a number of independent investigations (Gabriel 1992,
1995; Abrams \& Kumar 1996). The lineprofile for a p-mode of
frequency 1.9 mHz and $\ell=1$, calculated using JC-D solar model, is 
shown in figure 5.

The physical explanation for line asymmetry is simplest when sources lie
in the region of the sun where acoustic waves can propagate
(this case does not seem to apply to solar oscillations however
which are excited by sources that lie in the evanescent region).
Consider a source lying close to the node of a p-mode. As the frequency
of acoustic waves is varied in the neighborhood of this p-mode frequency
the position of the node changes with respect to the source position.
Thus waves of frequencies lying symmetrically on the opposite side
of the p-mode frequency gets excited to different amplitudes making the
resultant power spectrum asymmetrical. It is clear from this rather
simple example that not all p-mode line profiles are expected to be
equally asymmetrical (as is observed) and also that the degree of asymmetry
depends on the location of sources. In fact Duvall et al (1993) had
recognized this in their original paper and used this to determine the
depth of sources that are exciting p-modes. The recent paper of
Abrams and Kumar (1996) uses a realistic solar model due to
Christensen-Dalsgaard to calculate the p-mode power spectrum and finds 
that in order to reproduce the magnitude of asymmetry observed by 
Duvall et al. (1993) the sources responsible
for exciting low frequency p-modes should lie about
250 km below the photosphere. This might appear to be in conflict with the
result obtained using high frequency solar oscillations described in \S3.
However, the result is in agreement with the theory of stochastic excitation
which predicts that lower frequency oscillations are excited deeper in the
convection zone where the characteristic eddy time is longer.

Line asymmetry causes a slight error to the observational determination of 
p-mode frequencies which are obtained by fitting a Lorentzian function to 
the power spectra. Abrams and Kumar (1996) find that this frequency error
is proportional to the product of mode linewidth and a nondimensional 
measure of lineasymmetry $\eta_\alpha$ (see the figure below). The parameter
$\eta_\alpha$ is obtained by decomposing the observed power spectrum in the 
neighborhood of a peak corresponding to a
mode $\alpha$ into even and odd functions.  Since the odd function is zero
at the peak (by definition) and again far from the peak, its magnitude has 
a maximum at some intermediate distance from the peak, typically less than one
linewidth. The ratio of the maximum magnitude of the odd function to the
maximum magnitude of the even function is a dimensionless measure of the
asymmetry which we denote by $\eta_\alpha/100$ i.e. $\eta_\alpha$ is
the percentage line asymmetry of mode $\alpha$. The sign of $\eta_\alpha$
is taken to be positive or negative according to whether there is more
power on the high- or low-frequency side of the peak, respectively.

\begin{figure}[t]
\centerline{\psfig{figure=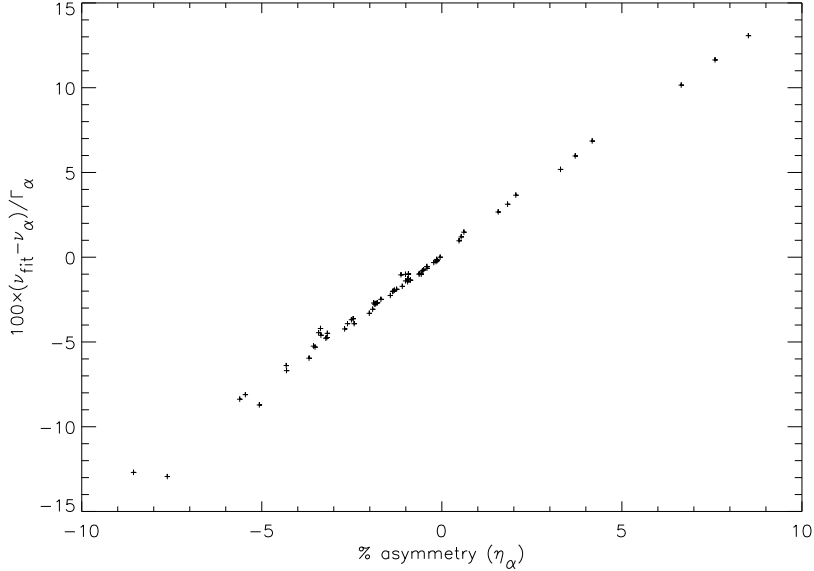,width=4.5in,angle=0.0}}
\caption[]{Error in the measurement of mode frequency (expressed as 
percentages of corresponding linewidths) as a result of asymmetry of lines 
in the power spectrum; $\ni_{fit}$ is the frequency obtained by fitting a 
Lorentzian function to the power spectrum, $\nu_\alpha$ is the mode 
eigenfrequency, and $\Gamma_\alpha$ is mode linewidth.
The frequency error is shown as a function of a dimensionless measure of
line asymmetry ($\eta_\alpha$) defined in the text. The slope of the line is 
approximately 1.5. The power spectra were calculated by solving an 
inhomogeneous
wave equation which included radiative damping of waves (see Abrams and Kumar 
1996, for details). The solar model used in this calculation was kindly 
provided by J. Christensen-Dalsgaard. 
}
\end{figure}

There is one feature of the observed line asymmetry that is very puzzling and
for which there is no theoretical explanation. Duvall et al. (1993) reported
that the sense of asymmetry reverses in the velocity and the intensity power 
spectra for the p-mode. This behavior has been confirmed by the most recent 
GONG data. The difference between the velocity and the intensity power spectra
can arise as a result of line formation in the presence of oscillations.
However, it is not clear what process can cause a reversal of the sign of 
$\eta_\alpha$ in the two spectra; the process has to be extremely
frequency sensitive so that it can modify the spectrum in an interval of 
only a few $\mu$Hz.

\section{Can we detect gravity modes in the sun?}

Gravity mode oscillations of the sun are primarily confined to its
radiative interior and their observation would thus provide a wealth
of information about the energy generating region which is poorly
probed by the p-modes. In the past 20 years a number of different
groups have claimed to detect g-modes in the sun (e.g. Brookes et al.
1976; Brown et al.  1978; Delache \& Scherrer 1983; Scherrer et al.
1979; Severny et al.  1976, Thomson et al. 1995; for a detailed review of 
the observations please see the article by Pall\'e, 1991, and references 
therein), but thusfar there is no consensus that g-modes have in fact been 
observed. One of the objectives of the instruments aboard SOHO (VIRGO, GOLF
and SOI) is to search for solar g-modes. So it should be helpful to 
estimate the expected surface velocity amplitudes of g-modes in the sun, 
providing observations and data analysis programs a rough target number 
to shoot for. The first question we need to address in this respect is
whether solar g-modes are self-excited (overstable) or not. This is dealt 
with in the following paragraph.

A number of people have investigated the linear 
stability of solar g-modes (e.g. Dilke \& Gough 1972; Rosenbluth \& Bahcall 
1973; Christensen-Dalsgaard et al. 1974; Shibahashi et al. 1975; Boury et
al. 1975; Saio 1980).  All of these investigations find that g-modes
of radial-order ($n$) greater than 3 are stable. However, there is no
general agreement about the stability of low order modes ($n\le 3$).
If overstable, the g-mode amplitude
will increase exponentially with time until nonlinear effects become
important and saturate their growth. Kumar and Goodman (1995) have
recently investigated 3-mode parametric interaction, a very efficient
nonlinear process.  Using their results we find that the low order
overstable g-modes in the sun will attain an energy of at least
10$^{37}$ erg before they are limited by nonlinearities. The velocity
at the solar surface corresponding to this energy is $\sim 10^2$ cm
s$^{-1}$, which is an order of magnitude larger than the observational
limit of Pall\'e (1991). Thus even low order g-modes of frequency
greater than about 150 $\mu$Hz are unlikely to be overstable.

\begin{figure}[b]
\centerline{\psfig{figure=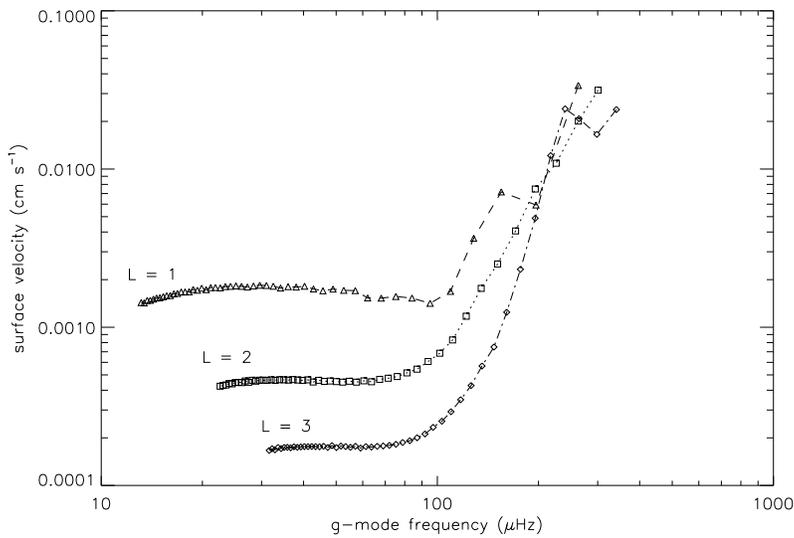,width=4.2in,angle=0.0}}
\caption[]{Magnitude of the surface velocity amplitude as a 
function of frequency for low degree solar g-modes excited by coupling 
with turbulent convection.  The surface velocity amplitude falls off
rapidly with increasing $\ell$, thus only low degree g-modes are
expected to be observable.}  
\end{figure}

However, g-modes can be stochastically excited. A number of people have
estimated g-mode amplitude assuming that they are linearly
stable and stochastically excited. Keeley (1980) applied his theory,
depeloped with Goldreich in 1977, of the excitation of solar modes
to estimate the amplitude of the 160 minute oscillation, and found the
theoretical amplitude to be much smaller than claimed by the observations;
much more sensitive observational searches since then have not detected
this oscillation (cf. Pall\'e 1991). Gough (1985) carried out an 
application of the energy partition result of Goldreich and Keeely (1977) 
to solar g-modes and estimated the surface velocity amplitude of low $n$ 
and $\ell$ g-modes ($n\le 3$, $\ell\le 2$) to be about 1-2 mm s$^{-1}$.
Kumar et al. (1996) estimated the g-mode amplitude using the
recent theoretical work of Goldreich et al. (1994) on
stochastic excitation of waves, which reproduces
the observed energy input rate into solar p-modes of all frequencies
(see \S 2), and taking into account the radiative and viscous turbulent
dampings. They find the surface velocity amplitude of low order g-modes to 
be about 0.4 mm s$^{-1}$ (see figure 7). Recently Anderson has carried
out numerical simulation of g-mode excitation as a result of
turbulent flow associated with penetrative convection. He finds that the
transverse surface velocity amplitudes of g-modes of degree about 6 is
$\sim$ 0.2 mm s$^{-1}$ in the case when he assumes that 10$^3$ modes are 
excited by this process (Anderson, 1996). Thus several different 
calculations suggest that the amplitudes for low degree g-modes are
of order 0.5 mm s$^{-1}$. The uncertainty in this estimate is at least a 
factor of a few. If the nature turns out to be cooperative and the
actual amplitudes of solar g-modes are a factor of a few larger than these
estimates, then instruments aboard SOHO have a good chance of detecting 
g-modes and opening up a new window in the study of the solar core.

\section*{Acknowledgment}

I am grateful to Peter Goldreich for collaboration over a number of
years and for his insights on the topic discussed in this review.
I thank Joergen Christensen-Dalsgaard for the use of his solar model,
and for pointing out several references. I am indebted to Tom Duvall,
Jack Harvey and Stuart Jefferies for sharing with me their excellent
data from the South Pole. This work was supported by a NASA grant 
NAGW-3936.

\section*{References}

\ni Abrams, D., \& Kumar, P., 1996, ApJ, to appear in Nov.

\ni Anderson, B.N., 1996, AA 312, 610

\refind Ando, H. and Osaki, Y. 1975, {\it Publ. Ast. Soc. Japan}, {\bf 27}, 581

\refind Antia, H. M., Chitre, S. M. and Narasimha, D. 1982, {\it Solar 
Physics},
{\bf 77}, 303

\refind Antia, H. M. Chitre, S. M. and Gough, D. O. 1988, {\it Advances in
Helio- and Asteroseismology}, IAU Symposium 123 (eds J. Christensen-Dalsgaard
and S. Frandsen) p. 371

\ni Balmforth, N.J., 1992, MNRAS 255, 603

\ni Balmforth, N.J., and Gough, D.O., 1990, Sol. Ph. 128, 161

\refind Bogdan, T.J., Hindman, B.W., Cally, P.S., \& Charbonneau, P., 1996,
ApJ 465, 406

\refind Boury, A., Gabriel, M., Noels, A., Scuflaire, R., \& Ledoux, P.
1975, AA, 41, 279

\ni Brown, T.M., Stebbins, R.T., and Hill, H.A. 1978, ApJ 223, 324

\ni Brookes, J.R., Isaak, G.R., and van der Raay, H.B. 1976, Nature 259, 92-95

\refind Christensen-Dalsgaard, J., Dilke, J. F. W. W., \& Gough, D. O. 1974,
M.N.R.A.S., 169, 429

\refind Chaplin, W.J., Elsworth, Y., Isaak, G.R., McLeod, C.P., Miller, B.A.
and New, R., 1996, IAU symposium no. 181 

\ni Crighton, D. G. 1975, {\it Prog. Aerospace Sci.}, {\bf 16}. 31

\ni Delache, P., and Scherrer, P.H. 1983, Nature 306, 651

\ni Dilke, J. F. W. W., \& Gough, D. O. 1972, Nature, 240, 262

\ni Christensen-Dalsgaard, J., \& Frandsen, S., 1983, Solar Phys., 82, 165

\refind Christensen-Dalsgaard, J., Gough, D.O., and Libbrecht, K. G., 1989,
   ApJ 341, L103
   
\refind Elsworth, Y, et al., 1995, in Fourth SOHO Workshop in Helioseismology,
ed. J.T. Hoeksema, V. Domingo, B. Fleck \& B. Battrick 

\refind Duvall, T.L.Jr., Jefferies, S.M., Harvey, J.W.,  Osaki, Y.,
and Pomerantz, M.A. 1993, ApJ, 410, 829

\refind Duvall, T.L., Jefferies, S.M., Harvey, J.W., \& Pomerantz, M.A.,
1993, Nature 362, 430 

\refind Duvall, T. L., Harvey, J. W., and Pomerantz, M. A. 1988, in 
{\it Advances in Helio- and Asteroseismology}, IAU Symposium 123, eds. 
J. Christensen- Dalsgaard and S. Frandsen, p. 37

\refind Elsworth, Y., Isaak, G. R., Jefferies, S. M., McLeod, C. P.,
New, R., Pall\'e, P. L., Regulo, C. Roca Cort\'es, T. 1990, MNRAS 242, 135

\ni Gabriel, M. 1992, AA, 265, 771

\ni Gabriel, M. 1995, AA, 299, 245

\ni Goldreich, P. and Keeley, D. A. 1977a, ApJ 211, 934

\ni Goldreich, P. and Keeley, D.K., 1977b, ApJ 212, 243

\ni Goldreich, P. \& Kumar, P., 1990, ApJ 363, 694

\ni Goldreich, P. \& Kumar, P., 1991, ApJ 374, 366

\ni Goldreich, P., \& Murray, N., 1994, ApJ 424, 480

\ni Goldreich, P., Murray, N., \& Kumar, P. 1994, ApJ, 424, 466

\refind Gough, D.O., 1977, ApJ 214, 196

\refind Gough, D.O., 1985, {\it Future missions in solar, heliospheric
and space plasma physics}, eds. E.J. Rolfe and B. Battrick, ESA SP-235, 183

\ni Jefferies, S.M., 1996, personal communication

\ni Kulsrud, R. M., 1955, ApJ 121, 461

\ni Leibacher, J. \& Stein, R.F., 1971, Astrophysics Lett., 7, 191

\ni Libbrecht, K.G., ApJ 334, 510

\ni Libbrecht, K.G., \& Woodard, M.F., 1991, Science 253, 152

\ni Lighthill, M.J., 1952, Proc. Roy. Soc. A211, 564

\refind Keeley, D.A., 1980, in {\it Nonradial and Nonlinear Stellar Pulsation},
(eds H. A. Hill and W. A. Dziembowski), Springer, Berlin p. 245

\ni Kumar, P., \& Lu, E., 1991, ApJ, 375, L35

\ni Kumar, P., 1994 ApJ, 428, 827

\refind  Kumar, P, Fardal, M.A., Jefferies, S.M., Duvall, T.L. Jr.,
Harvey J.W., and Pomerantz, M.A. 1994, ApJ, 422, L29

\ni Kumar, P., Franklin, J., \& Goldreich, P., 1988, ApJ 328, 879

\ni Kumar, P. and Goldreich, P., 1989, ApJ 342, 558

\ni Kumar, P. and Goldreich, P., and Kerswell, R., ApJ 427, 483

\ni Kumar, P., and Goodman, J. 1996, ApJ 466, 946

\ni Kumar, P., Quataert, E. J., and Bahcall, J.N., 1996, ApJ 458, L83

\refind Murray, N., 1993, in {\it Seismic Investigation of the Sun and Stars},
ASP conference series, vol. 42, ed. T. Brown, p3

\ni Pall\'e, P. L. 1991, Adv. Space Res., 11, 4, 29

\ni Rosenbluth, M., \& Bahcall, J. N.  1973, ApJ, 184, 9

\ni Saio, H. 1980, ApJ, 240, 685

\refind Scherrer, P.H., Wilcox, J.M., Kotov, V.A., Severny, A.B., and
Tsap, T.T. 1979, Nature 277, 635

\ni Severny, A.B., Kotov, V.A., and Tsap, T.T. 1976, Nature 259, 87

\refind Shibahashi, H., Osaki, Y., \& Unno, W. 1975, Publ. Astron. Soc.
Japan, 27, 401

\ni Stein, R.F., 1967, Solar Phys. 2, 385

\refind Stein, R.F., \& Nordlund, A. 1991, in Challenges to Theories of the
Structure of Moderate-Mass Stars, ed. D. Gough \& J. Toomre (Berlin: 
Springer-Verlag), 195

\refind Thomson, David J., Maclennan, Carol G., Lanzerotti, Louis J., 1995,
Nature, 376, 139

\ni Tomczyk, S., 1996, personal communication

\ni Toutain, T., \& Frohlich, C., 1992, AA, 257, 287 

\ni Ulrich R. K. 1970, {\it Ap. J.}, {\bf 162}, 993
   
\ni Vernazza, J.~E., Avrett, E.~H., \& Loeser, R. 1981, ApJS, 45, 635

\ni Wolf, C.L., 1972, ApJ 177, L87

\end{document}